# KALMAN FILTERED COMPRESSED SENSING

*Namrata Vaswani*

Dept. of Electrical and Computer Engineering, Iowa State University, Ames, IA, namrata@iastate.edu

**ABSTRACT**

We consider the problem of reconstructing time sequences of spatially *sparse* signals (with unknown and time-varying sparsity patterns) from a *limited number* of linear *"incoherent"* measurements, *in real-time*. The signals are sparse in some transform domain referred to as the sparsity basis. For a single spatial signal, the solution is provided by Compressed Sensing (CS). The question that we address is, for a sequence of sparse signals, can we do better than CS, if (a) the sparsity pattern of the signal's transform coefficients' vector changes slowly over time, and (b) a simple prior model on the temporal dynamics of its current non-zero elements is available. The overall idea of our solution is to use CS to estimate the support set of the initial signal's transform vector. At future times, run a reduced order Kalman filter with the currently estimated support and *estimate new additions to the support set by applying CS to the Kalman innovations or filtering error (whenever it is "large")*.

**Index Terms/Keywords:** compressed sensing, Kalman filtering, compressive sampling, sequential MMSE estimation

## 1. INTRODUCTION

We consider the problem of reconstructing time sequences of spatially *sparse* signals (with unknown and time-varying sparsity patterns) from a *limited number* of linear *"incoherent"* measurements, *in real-time*. The signals are sparse in some transform domain referred to as the "sparsity basis" [1]. A common example of such a problem is dynamic MRI or CT to image deforming human organs such as the beating heart or to image brain neural activation patterns (in response to stimuli) using fMRI. The ability to perform real-time MRI capture and reconstruction can make interventional MR practical [2]. Human organ images are usually piecewise smooth and thus the wavelet transform is a valid sparsity basis [1, 3]. Due to strong temporal dependencies, the sparsity pattern usually changes *slowly* over time. MRI captures a small (sub-Nyquist) number of Fourier transform coefficients of the image, which are known to be "incoherent" with respect to the wavelet transform [1, 3]. Other examples include sequentially estimating optical flow of a single deforming object (sparse in Fourier domain) from a set of randomly spaced optical flow measurements (e.g. those at high intensity variation points [4]), or real-time video reconstruction using single-pixel camera [5].

The solution to the static version of the above problem is provided by Compressed Sensing (CS) [1, 6, 7]. The noise-free observations case [1] is exact with high probability (w.h.p.) while the noisy case [7] has a small error w.h.p.. But existing solutions for the dynamic problem [5, 8] treat the entire time sequence as a single spatiotemporal signal and perform CS to reconstruct it. This is a batch solution (need to wait to get the entire observation sequence) and has very high complexity. An alternative would be to apply CS at each time separately, which is online and low-complexity, but will require many more measurements to achieve low error (see Fig. 1). The question that we address is: *can we do better than performing CS at each time separately, if (a) the sparsity pattern (support set) of the transform coefficients' vector changes slowly, i.e. every time, none or only a few elements of the support change, and (b) a simple prior model on the temporal dynamics of its current non-zero elements is available (but do not know which coordinates are non-zero)*.

Our solution is motivated by reformulating the above problem as causal minimum mean squared error (MMSE) estimation with a slow time-varying set of dominant basis directions (or equivalently the support of the transform vector). If the support is known, the MMSE solution is given by the Kalman filter (KF) [9] for this support. But what happens if the support is unknown and time-varying? The initial support can be estimated using CS [7]. If at a given time, there is an addition to the support set, but we run the KF for the old model, there will be a model mismatch and the innovation (and filtering) error will increase. Whenever it does, *the change in support can be estimated by running CS on the innovation or the filtering error*, followed by thresholding. A Kalman update step is run using the new support set. If some coefficients become and remain nearly zero (or nearly constant), they can be removed from the support set.

If, for a moment, we assume that CS [7] gives the correct estimate of the support at all times, then the above approach will give the MMSE estimate of the signal at all times. The reason it is very likely that CS [7] gives the correct estimate is because we use it to fit a very sparse "model change" signal to the filtering error. Also note that a full Kalman filter [9], that does not use the fact that the signal is sparse, is meaningless here, because the number of observations available is smaller than the signal dimension, and thus many elements of the signal transform will be unobservable. Unless all unobservable modes are stable, the error will blow up. Other recent work that also attempts to use prior knowledge with CS, but to reconstruct only a single signal is [10, 11, 12].

## 2. THE MODEL AND PROBLEM FORMULATION

Let $(z_t)_{m \times 1}$ denote the spatial signal of interest at time $t$ and $(y_t)_{n \times 1}$, with $n < m$, denote its observation vector at $t$. The signal, $z_t$, is sparse in a given sparsity basis (e.g. wavelet) with orthonormal basis matrix, $\Phi_{m \times m}$, i.e. $x_t \triangleq \Phi' z_t$ is a sparse vector (only $S_t << m$ elements of $x_t$ are non-zero). Here ' denotes transpose. The observations are "incoherent" w.r.t. the sparsity basis of the signal, i.e. $y_t = Hz_t + w_t = H\Phi x_t + w_t$, where $H_{n \times m}$ is such that the correlation between the columns of $A \triangleq H\Phi$ is small enough to ensure that any $S_t$-column sub-matrix of $A$ is "approximately orthonormal" (its nonzero singular values are between $\sqrt{1-\delta}$ to $\sqrt{1+\delta}$ for $\delta < 1$). $w_t$ is the measurement noise. Thus the measurement model is:

$$y_t = Ax_t + w_t, \ A \triangleq H\Phi, \ w_t \sim \mathcal{N}(0, \sigma_{obs}^2 I) \qquad (1)$$

with $w_t$ being temporally i.i.d.. We refer to $x_t$ as the state at $t$. Our goal is to get the "best" causal estimate of $x_t$ (or equivalently of $z_t = \Phi x_t$) at each $t$.

Let $T_t$ denote the the support set of $x_t$, i.e. the set of its non-zero coordinates and let $S_t = size(T_t)$. In other words, $T_t = [i_1, i_2, \ldots i_{S_t}]$ where $i_k$ are the non-zero coordinates of $x_t$. For any set $T$, let $(v)_T$ denote the $size(T)$ length sub-vector containing the elements of $v$ corresponding to the indices in the set $T$. For another set, $\gamma$, we also use the notation $T_\gamma$ which treats $T$ as a vector and selects the elements of $T$ corresponding to the indices in the set $\gamma$. For a matrix $A$, $A_T$ denotes the sub-matrix obtained by extracting the columns of $A$ corresponding to the indices in $T$. We use the notation $(Q)_{T_1, T_2}$ to denote the sub-matrix of $Q$ containing rows and columns corresponding to the entries in $T_1$ and $T_2$ respectively. The set operations $\cup$, $\cap$, and $\setminus$ have the usual meanings (note $T_1 \setminus T_2$ denotes elements of $T_1$ not in $T_2$). We use $'$ to denote transpose. $T^c$ denotes the complement of $T$ w.r.t. $[1:m]$, i.e. $T^c \triangleq [1:m] \setminus T$. Also $||v||_p$ is the $l_p$ norm of the vector $v$, i.e. $||v||_p \triangleq (\sum_i |v_i|^p)^{1/p}$.

*Assumption 1.* We assume slow changes in sparsity patterns, i.e. the maximum size of the change in the support set at any time is smaller (usually much smaller) than $S_t$ at any $t$, i.e. $S_{diff,max} \triangleq \max_t[size(T_t \setminus T_{t-1}) + size(T_{t-1} \setminus T_t)] < \min_t S_t$. This, as we shall see later, ensures that performing CS in Gaussian noise to only estimate the additions to the support set results in smaller error w.h.p. than doing CS to estimate the entire $T_t$ again.

*Assumption 2.* We also assume that $A$ satisfies the Uniform Uncertainty Principle (UUP) (equation 1.6 of [7]) at a sparsity level, $S_{max} \triangleq \max_t S_t$. At first thought, it may appear that having $A$ satisfy UUP at the smaller level $S_{diff,max}$ is sufficient, but, it is not. This is because if $S_t > S_{diff,max}$ for some $t$, even though $A_{T_t}$ is a tall matrix, there is no guarantee that at least $S_t$ of its columns are linearly independent, i.e. it may happen that $rank(A_{T_t}) < S_t$. In this case, the system no longer remains observable for some elements of $x_t$. If any of these elements follow an unstable prior dynamic model, the KF error will increase unboundedly with $t$.

*System Model for $x_t$.* For the currently non-zero coefficients of $x_t$, we assume a spatially i.i.d. Gaussian random walk model, with noise variance $\sigma_{sys}^2$. At the first time instant at which $(x_t)_i$ becomes non-zero, it is assumed to be generated from a zero mean Gaussian with variance $\sigma_{init}^2$. Thus, we have the model: $x_0 = 0$,

$(x_t)_i = (x_{t-1})_i + (\nu_t)_i, \ (\nu_t)_i \sim \mathcal{N}(0, \sigma_{sys}^2), \ \text{if } i \in T_t, \ i \in T_{t-1}$
$(x_t)_i = (x_{t-1})_i + (\nu_t)_i, \ (\nu_t)_i \sim \mathcal{N}(0, \sigma_{init}^2) \text{ if } i \in T_t, \ i \notin T_{t-1}$
$(x_t)_i = (x_{t-1})_i \text{ if } i \notin T_t \quad (2)$

The above model can be compactly written as: $x_0 = 0$,

$$x_t = x_{t-1} + \nu_t, \ \nu_t \sim \mathcal{N}(0, Q_t),$$
$$(Q_t)_{T_t \cap T_{t-1}, T_t \cap T_{t-1}} = \sigma_{sys}^2 I$$
$$(Q_t)_{T_t \setminus T_{t-1}, T_t \setminus T_{t-1}} = \sigma_{init}^2 I$$
$$(Q_t)_{T_t^c, T_t^c} = 0 \quad (3)$$

where *the set $T_t$ is unknown* $\forall t$. If $T_t$ were known at each $t$, i.e. the system model was completely defined, the MMSE estimate of $x_t$ from $y_1, y_2, \ldots y_t$ would be given by a reduced order KF defined for $(x_t)_{T_t}$. But, as explained in Sec. 1, in most practical problems, $T_t$ is in fact unknown and time-varying. Often, it may be possible to get a rough prior estimate of $T_1$ by thresholding the eigenvalues of the covariance of $x_1$ (possible to do if multiple realizations of $x_1$ are available to estimate its covariance). But without multiple i.i.d. realizations of the entire $\{x_t\}$, which are impossible to obtain in most cases, it is not possible to get a-priori estimates of $T_t$ for all $t$. But note that, it is possible to estimate $\sigma_{sys}^2, \sigma_{init}^2$ for the model of (3) using just one "training" realization of $\{x_t\}$ (which is usually easy to get) by setting the near-zero elements to zero in each $x_t$ and using the rest to obtain an ML estimate.

Assuming known values of $\sigma_{sys}^2, \sigma_{init}^2$, our goal here is to get the best estimates of $T_t$ and $x_t$ at each $t$ using $y_1, \ldots y_t$. Specifically,

1. At each time, $t$, get the best estimate of the support set, $T_t$, i.e. get an estimate $\hat{T}_t$ with smallest possible $[size(\hat{T}_t \setminus T_t) + size(\hat{T}_t \setminus T_t)]$ using $y_1, y_2 \ldots y_t$.

2. Assuming the estimates of $T_1, \ldots T_t$ are perfect (have zero error), get the MMSE estimate of $x_t$ using $y_1, y_2 \ldots y_t$.

## 3. KALMAN FILTERED COMPRESSED SENSING (KF-CS)

We explain the proposed KF-CS algorithm below.

*Running the KF.* Assume, for now, that the support set at the first time instant, $T_1$, is known. Consider the situation where the first change in the support occurs at a $t = t_a$, i.e. for $t < t_a$, $T_t = T_1$, and that the change is an addition to the support. This means that for $t < t_a$, we need to just run a regular KF, which assumes the following reduced order measurement and system models: $y_t = A_T(x_t)_T + w_t$, $(x_t)_T = (x_{t-1})_T + (\nu_t)_T$, with $T = T_1$. The KF prediction step for this model is [9]:

$$\hat{x}_{t|t-1} = \hat{x}_{t-1}$$
$$(P_{t|t-1})_{T,T} = (P_{t-1})_{T,T} + \sigma_{sys}^2 I \quad (4)$$

with initialization $\hat{x}_0 = 0$, $P_0 = 0$. The update step is [9]:

$$K \triangleq (P_{t|t-1})_{T,T} A'_T \Sigma_{ie}^{-1}, \ \Sigma_{ie} \triangleq A_T (P_{t|t-1})_{T,T} A'_T + \sigma_{obs}^2 I$$
$$(\hat{x}_t)_T = (\hat{x}_{t|t-1})_T + K[y_t - A\hat{x}_{t|t-1}]$$
$$(\hat{x}_t)_{T^c} = (\hat{x}_{t|t-1})_{T^c} = (\hat{x}_{t-1})_{T^c}$$
$$(P_t)_{T,T} = [I - KA_T](P_{t|t-1})_{T,T} \quad (5)$$

Note the update step for the elements of $T^c$.

*Detecting If Addition to Support Set Occurred.* The Kalman innovation error is $\tilde{y}_t \triangleq y_t - A\hat{x}_{t|t-1}$. For $t < t_a$, $\tilde{y}_t = A(x_t - \hat{x}_{t|t-1}) + w_t \sim \mathcal{N}(0, \Sigma_{ie})$ [9]. At $t = t_a$, a new set, $\Delta$, gets added to the support of $x_t$, i.e. $y_t = A_T(x_t)_T + A_\Delta(x_t)_\Delta + w_t$, where the set $\Delta$ is unknown. Since the old model is used for the KF prediction, $\tilde{y}_t$ at $t = t_a$ will have non-zero mean, $A_\Delta(x_t)_\Delta$, i.e. at $t = t_a$,

$$\tilde{y}_t = A_\Delta(x_t)_\Delta + \tilde{w}_t, \ \tilde{w}_t \triangleq [A_T(x_t - \hat{x}_{t|t-1})_T + w_t] \sim \mathcal{N}(0, \Sigma_{ie}) \quad (6)$$

Thus, the problem of detecting if a new set has been added or not gets transformed into the problem of detecting if the Gaussian distributed $\tilde{y}_t$ has non-zero or zero mean. We use the well-known generalized Likelihood Ratio Test for this problem, which detects if the weighted innovation error norm, $IEN \triangleq \tilde{y}'_t \Sigma_{ie}^{-1} \tilde{y}_t \gtrless$ threshold. One can replace $IEN$ by the weighted norm of the filtering error, defined below, which will make the detection more sensitive.

*Estimating the Additions (using CS).* If the $IEN$ (or $FEN$) is "high" (above the detection threshold), there is a need to estimate $\Delta$. At $t = t_a$, the innovation error, $\tilde{y}_t$, can also be written as:

$$\tilde{y}_t = A_{T^c}(x_t)_{T^c} + \tilde{w}_t, \ \tilde{w}_t \sim \mathcal{N}(0, \Sigma_{ie}) \quad (7)$$

while the filtering error, $\tilde{y}_{t,f} \triangleq y_t - A\hat{x}_t$, can be written as:

$$\tilde{y}_{t,f} = A_{T^c}(x_t)_{T^c} + A_T(x_t - \hat{x}_t)_T + w_t$$
$$= [I - A_T K] A_{T^c}(x_t)_{T^c} + \tilde{w}_{t,f}, \ \tilde{w}_{t,f} \triangleq [I - A_T K]\tilde{w}_t$$
$$\tilde{w}_{t,f} \sim \mathcal{N}(0, \Sigma_{fe}), \ \Sigma_{fe} \triangleq [I - A_T K]\Sigma_{ie}[I - A_T K]' \quad (8)$$

**Algorithm 1** Kalman Filtered Compressive Sensing (KF-CS)

Initialization: Set $\hat{x}_0 = 0$, $P_0 = 0$, $T_0$= empty (if unknown) or equal to the known/partially known support. For $t > 0$, do,

1. Set $T \leftarrow T_{t-1}$.
2. **KF prediction.** Run (4) using the current $T$.
3. **KF update.** Run (5) using the current $T$.
4. **Addition (using CS).** Compute $IEN = \tilde{y}_t' \Sigma_{ie}^{-1} \tilde{y}_t$ where $\tilde{y}_t \triangleq y_t - A\hat{x}_{t|t-1}$ (or compute $FEN = \tilde{y}_{t,f}' \Sigma_{fe}^{-1} \tilde{y}_{t,f}$ where $\tilde{y}_{t,f} \triangleq y_t - A\hat{x}_t$), and check if it is greater than its threshold. If it is,
    (a) **Run CS on the filtering error**, $\tilde{y}_{t,f} \triangleq y_t - A\hat{x}_t$, i.e. compute the Dantzig selector using (9).
    (b) Compute the support set of $x_{new}$ by thresholding, i.e. compute $nz = \{i : |(x_{new})_i| > \alpha\}$. Then the addition to the support set of $x_t$ is $\Delta = (T^c)_{nz}$. The new support set for $x_t$ is $T_{new} = T \cup \Delta$.
    (c) Set $(P_{t|t-1})_{\Delta,\Delta} = \sigma_{init}^2 I$. Set $T \leftarrow T_{new}$.
    (d) **Run the KF update** given in (5) for the current $T$.

   Performance can be improved by iterating the above four steps until $size(\Delta) = 0$ or $FEN$ less than its threshold.
5. **Deletion.** Compute the set $\Delta_D = \{i \in T : \sum_{\tau=t-k+1}^{t}(\hat{x}_\tau)_i^2 < k\alpha^2\}$. The new support set is $T_{new} = T \setminus \Delta_D$.
    (a) Set $(P_{t|t-1})_{\Delta_D,[1:m]} = 0$, $(P_{t|t-1})_{[1:m],\Delta_D} = 0$. Set $T \leftarrow T_{new}$.
6. **KF update.** Run (5) for the current $T$.
7. Assign $T_t \leftarrow T$. **Output $T_t$, $\hat{x}_t$ and the signal estimate, $\hat{z}_t = \Phi \hat{x}_t$.** Increment $t$ and go to step 1.

---

with $(x_t)_{T^c}$ being a sparse vector with support, $nz$, s.t. $\Delta = (T^c)_{nz}$. Note that $\Sigma_{fe} < \Sigma_{ie}$. The problem of estimating $(x_t)_{T^c}$ from either $\tilde{y}_t$ or $\tilde{y}_{t,f}$ is the problem of compressed sensing in Gaussian noise studied in [7], except that the noise now is spatially colored. It is not immediately clear whether to use $\tilde{y}_t$ or $\tilde{y}_{t,f}$. In $\tilde{y}_{t,f}$, the "noise", $\tilde{w}_{t,f}$, is smaller (the change $(x_t - x_{t-1})_T$ has been estimated and subtracted out), but the new component, $A_{T^c}(x_t)_{T^c}$, is also partially suppressed. But the suppression is small because $A_T K A_{T^c}(x_t)_{T^c} = A_T(P_{t|t-1}^{-1}\sigma_{obs}^2 + A_T'A_T)^{-1}A_T'A_{T^c}(x_t)_{T^c}$ (follows by rewriting $K$ using matrix inversion lemma [9]) and $A_T'A_{T^c}(x_t)_{T^c} = A_T'A_\Delta(x_t)_\Delta$ is small (since $A$ satisfies UUP at level $S_{max}$ and $S_{t_a} = size(T \cup \Delta) \leq S_{max}$). Thus, we use $\tilde{y}_{t,f}$.

The Dantzig selector [7] can be applied to $\tilde{y}_{t,f}$ to estimate $(x_t)_{T^c}$, followed by using thresholding to compute its support, $nz$, and thus get $\Delta$. Since $\tilde{w}_{t,f}$ is colored, one way to modify the Dantzig selector to apply it to (8) is to compute:

$$x_{new} = \arg\min_\beta ||\beta||_1, \text{s.t.} ||A_{T^c}' \Lambda^{-1/2} U'[\tilde{y}_{t,f} - A_{T^c}\beta]||_\infty \leq \lambda_m \quad (9)$$

where $\Sigma_{fe} = U\Lambda U'$ is the eigenvalue decomposition of $\Sigma_{fe}$ and $\lambda_m \triangleq \sqrt{2\log m} \ \text{trace}(\Sigma_{fe})/n$. The set $nz$ is estimated as $nz = \{i : |(x_{new})_i| > \alpha\}$ for some zeroing threshold $\alpha$ and the set $\Delta$ is $\Delta = (T^c)_{nz}$. Thus the new support set is $T_{new} = T \cup \Delta$. We initialize the prediction covariance along $\Delta$: $(P_{t|t-1})_{\Delta,\Delta} = \sigma_{init}^2 I$.

*KF Update.* We run the KF update given in (5) with $T = T_{new}$. This can be interpreted as a Bayesian version of Gauss-Dantzig [7].

*Iterating CS and KF-update.* Often, it may happen that not all the elements of the true $\Delta$ get estimated in one run of the CS step, because of the error in CS. To address this, the above steps (CS and KF update) can be iterated until $FEN$ goes below a threshold or until the estimated set $\Delta$ is empty or for a fixed number of iterations. A similar idea forms the basis of iterative CS reconstruction techniques such as stagewise Orthogonal Matching Pursuit[13].

*Deleting Near-Zero Coefficients.* Over time, some coefficients may become zero (or nearly-zero) and remain zero. Alternatively, some coefficients may wrongly get added, due to CS error. In both cases, the coefficients need to be removed from the support set $T_t$. A simple way to do this would be to check if $|(\hat{x}_t)_i| < \alpha$ ($\alpha$ is a zeroing threshold) for a few time instants. When a coefficient is removed, the corresponding row and column in $P_{t|t-1}$ is set to zero.

*Deleting Constant Coefficients.* If a coefficient, $i$, becomes nearly-constant (this may happen in certain applications), one can keep improving the estimate of its constant value by changing the prediction step for it to $(P_{t|t-1})_{i,i} = (P_{t-1})_{i,i}$. Either one can keep doing this forever (the error in its estimate will go to zero with $t$) or one can assume that the estimation error has become negligibly small after a finite time and then remove the coefficient index from $T_t$. The bias-variance tradeoff needs to be evaluated before selecting either scheme. If the latter is done, then for future times, one needs to replace $y_t$ by $y_t - A_i(\hat{x})_i$ and set $(P_{t|t-1})_{i,[1:m]} = 0$, $(P_{t|t-1})_{[1:m],i} = 0$. This will be part of future work.

*Initialization.* Initially, the support set, $T_1$ may be roughly known (estimated by thresholding the eigenvalues of the covariance of $x_1$, which is computable if its multiple realizations are available) or unknown. We initialize KF-CS by setting $\hat{x}_0 = 0$, $P_0 = 0$ and $T_0 =$ roughly known support or $T_0 =$ empty (if support is completely unknown). In the latter case, automatically at $t = 1$, the $IEN$ (or $FEN$) will be large, and thus CS will run to estimate $T_1$.

The entire KF-CS algorithm is summarized in Algorithm 1.

### 3.1. Discussion

The key difference between KF-CS and regular CS at each $t$ is that KF-CS performs *CS on the filtering error, $\tilde{y}_{t,f}$, to only detect new additions* while regular CS performs CS on the observation, $y_t$, to detect the entire vector $x_t$ (without using knowledge of the previous support set). From [7, Theorem 1.1] (the result should hold directly or with slight modification for CS in colored noise, but we have not checked it yet), the CS error is directly proportional to the sparsity size of the vector being estimated and to the noise variance. But the dependence on sparsity size is much stronger (highly non-linear) while that on the noise variance is linear. Note that "CS on filtering error" needs to estimate at most $S_{diff,\max}$ coefficients and $S_{diff,\max} < S_t$, $\forall t$ (Assumption 1). Thus, if the "noises" in both cases were the same, then the error in "CS on filtering error" would be much smaller than that in CS. But the "noise" in $y_t$ is only $w_t$, while the "noise" in $\tilde{y}_{t,f}$ is $w_t$ plus $A_T(x_t - \hat{x}_t)_T$. If we can show stability of KF-CS error, and if the temporal prior is strong enough, the covariance of this extra "noise" will be small enough. Thus, it should be possible to show that the error in KF-CS is smaller than that in regular CS, if the sparsity pattern changes slowly enough and the temporal prior is strong enough.

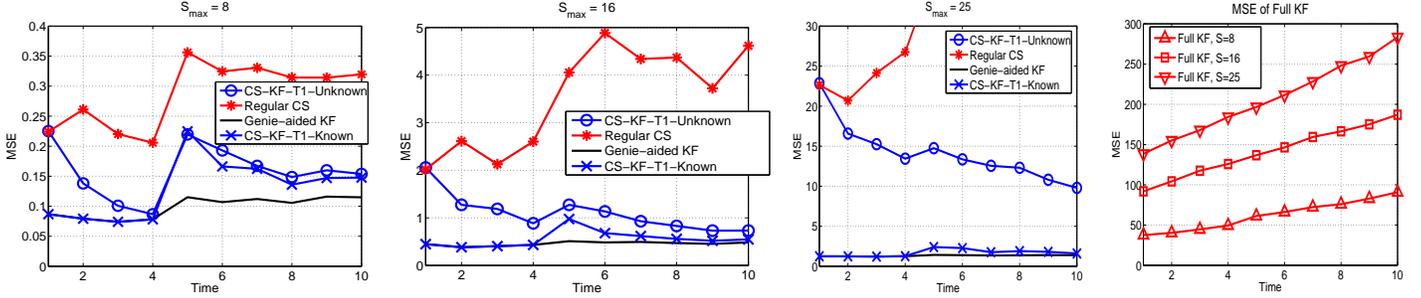

**Fig. 1**. MSE plots of KF-CS (labeled CS-KF in the plots), $T_1$ unknown and known cases, compared against regular CS in the first 3 figures and against the Full 256-dim KF in the last figure (its MSE is so large that we cannot plot it in the same scale as the others). The benchmark (MMSE estimate with known $T_1, T_5$) is the genie-aided KF. The simulated signal's energy at $t$ is $\mathbb{E}[||x_t||_2^2] = S_1\sigma_{init}^2 + (\sum_{\tau=2}^{t} S_\tau)\sigma_{sys}^2$.

A rigorous comparison of KF-CS with CS and an analysis of KF-CS error stability will be part of future work. We provide here a qualitative discussion of the sources of error in KF-CS. The error can be due to two reasons - either the CS step misses some non-zero coefficients (or the deletion step wrongly deletes a non-zero coefficient) or the CS step estimates some extra coefficients (or the deletion step misses removing some coefficients).

Occasionally, an element, $i$, of the true $\Delta$ may get missed, due to nonzero CS error (or the thresholding error). The extra error due to an $(x_t)_i$ getting missed cannot be larger than the CS error at the current time (which itself is upper bounded by a small value w.h.p. [7]) plus $\alpha$ (due to thresholding). Also, eventually, when the magnitude of $(x_t)_i$ increases at a future time, it will result in a "high" $IEN$ and $FEN$, and the CS step at that time will estimate it w.h.p..

We can prevent too many extra coordinates from getting wrongly estimated by having a rough idea of the maximum sparsity of $x_t$ and using thresholding to only select that many, or a few more, highest magnitude non-zero elements. The deletion threshold also needs to be selected appropriately. Also, if, because of CS thresholding or deletion, some true element gets missed because its value was too small, it will, w.h.p., get detected by CS at a future time when $IEN$ increases. Also, as long as $rank(A_T) > size(T)$ for the currently estimated $T$ (which may contain some extra coordinates), the estimation error will increase beyond MMSE, but will not blow up, i.e. it will still converge, but to a higher constant value.

## 4. SIMULATION RESULTS

We simulated a time sequence of sparse $m=256$ length signals, $x_t$, with maximum sparsity $S_{max}$. Three sets of simulations were run with $S_{max}$= 8, 16 and 25. The $A$ matrix was simulated as in [7] by generating $n \times m$ i.i.d. Gaussian entries (with $n = 72$) and normalizing each column of the resulting matrix. Such a matrix has been shown to satisfy the UUP at a level $C \log m$ [7]. The observation noise variance, $\sigma_{obs}^2 = ((1/3)\sqrt{S_{max}/n})^2$ (this is taken from [7]). The prior model on $x_t$ was (3) with $\sigma_{init}^2 = 9$ and $\sigma_{sys}^2 = 1$. $T_1$ (support set of $x_1$) was obtained by generating $S_{max} - 2$ unique indices uniformly randomly from $[1:m]$. We simulated an increase in the support at $t = 5$, i.e. $T_t = T_1$, $\forall t < 5$, while at $t = 5$, we added two more elements to the support set. Thus, $T_t = T_5$, $\forall t \geq 5$ had size $S_{max}$. Only addition to the support was simulated.

We used the proposed KF-CS algorithm (Algorithm 1) to compute the causal estimate $\hat{x}_t$ of $x_t$ at each $t$. The resulting mean squared error (MSE) at each $t$, $\mathbb{E}_{x,y}[||x_t - \hat{x}_t||_2^2]$, was computed by averaging over 100 Monte Carlo simulations of the above model. The same matrix, $A$, was used in all the simulations, but we averaged over the joint pdf of $x, y$, i.e. we generated $T_1, T_5, (\nu_t)_{T_t}, , w_t, t = 1, \ldots 10$ randomly in each simulation. Our simulation results are shown in Fig. 1 *(KF-CS is labeled as CS-KF in plots by mistake)*. Our benchmark was the genie-aided KF, i.e. an $S_{max}$-order KF with known $T_1$ and $T_5$, which generates the MMSE estimate of $x_t$. We simulated two types of KF-CS methods, one with known $T_1$, but unknown $T_5$ and the other with unknown $T_1$ and $T_5$. Both performed almost equally well for $S_{max} = 8$, but as $S_{max}$ was increased much beyond the UUP level of $A$, the performance of the unknown $T_1$ case degraded more (the CS assumption did not hold). We also show comparison with regular CS at each $t$, which does not use the fact that $T_t$ changes slowly (and does not assume known $T_1$ either). This had much higher MSE than KF-CS. The MSE become worse for larger $S_{max}$. We also implemented the full KF for the 256-dim state vector. This used (3) with $Q_t = \sigma_{sys}^2 I_{256 \times 256}$, i.e. it assumed no knowledge of the sparsity. Since we had only a 72-length observation vector, the full system is not observable. Since all non-zero modes are unstable, its error blows up.

## 5. CONCLUSIONS AND FUTURE DIRECTIONS

To the best of our knowledge, this is the first work on extending the CS idea to *causally* estimate a time sequence of spatially sparse signals. We do this by using CS to estimate the signal support at the initial time instant, followed by running a KF for the reduced order model, until the innovation or filtering error increases. When it does, we estimate the "change in support" by running CS on the filtering error. This has much lower error since the "change" is much sparser than the actual signal. Open questions to be addressed in future are (a) the analysis of the stability of KF-CS, (b) comparison of KF-CS error with that of regular CS, (c) studying how and when to delete constant coefficients, (d) KF-CS for compressible signal sequences, and (e) its extensions to large-dimensional particle filtering [14].